\documentstyle[12pt]{article}
\input epsf
\newcommand{\e}[1]{\label{eq:#1}}
\newcommand{\ee}[1]{(\ref{eq:#1})}

\newcommand{\eq}{\begin{equation}}
\newcommand{\eqe}{\end{equation}}
\newcommand{\eqa}{\begin{eqnarray}}
\newcommand{\eqae}{\end{eqnarray}}

\newcommand{\del}{\partial}
\newcommand{\slm}{ \rm }

\newcommand{\bq}{ {\bf q} }
\newcommand{\bk}{ {\bf k} }
\newcommand{\bl}{ {\bf l} }
\newcommand{\bp}{ {\bf p} }
\newcommand{\bD}{ {\bf D} }
\newcommand{\br}{ {\bf r} }
\newcommand{\bP}{ {\bf P} }
\newcommand{\bv}{ {\bf v} }
\begin{document}

\pagestyle{empty}
\hfill{NSF-ITP-92-132}

\hfill{UTTG-20-92}

\hfill{hep-th/9210046}

\vspace{24pt}

\begin{center}
{\bf Effective Field Theory and the Fermi Surface }

\vspace{24pt}
\renewcommand{\thefootnote}{\fnsymbol{footnote}}
Joseph Polchinski\footnote{joep@sbitp.ucsb.edu}
\renewcommand{\thefootnote}{\arabic{footnote}}
\addtocounter{footnote}{-1}

\vspace{24pt}
\sl 

Institute for Theoretical Physics \\ University of California \\
Santa Barbara, California 93106-4030
 
\vspace{12pt}
\rm and\\
\vspace{12pt}
\sl
Theory Group\\ Department of Physics \\ University of Texas
\\ Austin, Texas 78712
\rm

\vspace{24pt}
{\bf ABSTRACT}
 
\end{center}
\baselineskip=18pt
 
\begin{minipage}{4.8in}
This is an introduction to the method of effective field theory.  
As an application, I derive
the effective field theory of low energy excitations in a conductor, the
Landau theory of Fermi liquids, 
and explain why the high-$T_c$ superconductors 
must be described by a different effective field theory.\\[6pt]
Lectures presented at TASI 1992.
\end{minipage}
 
\vfill
 
\pagebreak
\baselineskip=18pt
\pagestyle{plain}
\setcounter{page}{1}

Effective field theory is a very powerful tool in quantum field theory,
and in particular gives a new point of view about the meaning of
renormalization.  It is approximately two decades old and appears 
throughout the
literature, but it is ignored in most textbooks.  It is therefore 
appropriate to devote two lectures to effective field theory here at the TASI
school.\footnote{See also the lecture by Peter Lepage in the 1989 TASI
Proceedings.}

In the first lecture I will describe the general method and ideology.  In the
second I will develop in detail one application---the effective field theory
of the low-energy excitations in a metal, which is known as the Landau theory
of Fermi liquids.  This is an unusual subject for a school on particle physics,
but you will see that it is a beautiful example of the main themes in effective
field theory.  As a bonus, we will be able to understand
something about the high temperature superconducting materials, and why
it appears that they require a new idea in quantum field theory. 

\section*{Lecture 1: Effective Field Theory}

Consider a quantum field theory with a characteristic energy scale $E_0$,
and suppose we are interested in 
the physics at some lower scale $E << E_0$.  Of
course, most systems have several characteristic scales, but we can consider
them one at a time.  The next lecture will illustrate the treatment
of a system with two scales.

Effective field theory is a general method for analyzing this situation.
Choose a cutoff $\Lambda$ at or slightly below $E_0$, and
divide the fields in the path integral into high- and low- frequency parts,
\eqa
\phi &=& \phi_{\slm H} + \phi_{\slm L} \nonumber\\[5pt]
&\phi_{\slm H}:& \omega > \Lambda  \nonumber\\[2pt]
&\phi_{\slm L}:& \omega < \Lambda. \e{momsep}
\eqae
For the rather general purposes of these lectures, we do not need to specify
how this is done---whether the separation is sharp or smooth, how Lorentz and
other symmetries are preserved, and so on.  Now do the path integral over the
high-frequency part,
\eq
\int {\cal D} \phi_{\slm L} \, {\cal D} \phi_{\slm H}\,
e^{i S(\phi_{\slm L}, \phi_{\slm H}) }
= \int {\cal D} \phi_{\slm L}\,
e^{i S_\Lambda (\phi_{\slm L}) },  \e{pathint}
\eqe
where
\eq
e^{i S_\Lambda (\phi_{\slm L}) }
= \int {\cal D} \phi_{\slm H}\,
e^{i S(\phi_{\slm L}, \phi_{\slm H}) }. \e{highint}
\eqe
We are left with an integral with an upper frequency cutoff $\Lambda$ and an
effective action $S_\Lambda (\phi_{\slm L})$.  This is known as a {\it low
energy} or {\it Wilsonian} effective action, to distinguish it from the 1PI
effective action generated by integrating over all frequencies but keeping only
1PI graphs.  This point of view is identified with Wilson,
though it has many roots in the literature; see the Bibliography.

We can expand $S_\Lambda$ in terms of local operators ${\cal
O}_i$,\footnote
{Because it is generated by integrating out modes $\omega > \Lambda$,
$S_\Lambda$ is nonlocal in time on the scale $1/\Lambda$.  
The expansion in local
operators is possible because the remaining fields are of frequency $\omega <
\Lambda$. One could keep the theory strictly local in time, at the cost of
Lorentz invariance, by imposing the cutoff in spatial momentum only.}
\eq
S_\Lambda = \int d^D x\, \sum_i g_i {\cal O}_i. \e{effact}
\eqe
The sum runs over all local operators allowed by the symmetries of the problem.
This is an infinite sum; to make this approach 
useful we now do some dimensional
analysis.  In units of $\hbar = 1$, the action is dimensionless; for the
purposes of the present section we also set $c = 1$.  Use the
free action to assign units to the fields (more about this later).
If an operator ${\cal O}_i$ has units $E^{\delta_i}$, then $\delta_i$Ê is known
as its dimension and $g_i$ has units $E^{D - \delta_i}$ where $D$ is the
spacetime dimension.  For example, in scalar field theory, the free action
\eq
\frac{1}{2} \int d^D x\, \del_\mu\phi \del^\mu \phi \e{scakin}
\eqe
determines the units of $\phi$ to be $E^{-1 + d/2}$.  An operator ${\cal O}_i$
constructed from $M$ $\phi$'s and $N$ derivatives then has dimension
\eq
\delta_i = M(-1 + D/2) + N. \e{scalopdim}
\eqe

Now define dimensionless couplings $\lambda_i = \Lambda^{\delta_i - D} g_i$.
Since $\Lambda$ is essentially the characteristic scale of the system, the
$\lambda_i$ are presumably of order~1.  Now, for a process at scale $E$,
we estimate dimensionally the magnitude of a given term in the action as
\eq
\int d^D x\, {\cal O}_i \sim E^{\delta_i - D}, \e{opest}
\eqe
so that the $i$'th term is of order
\eq
\lambda_i \left( \frac{E}{\Lambda} \right)^{\delta_i - D}.
\eqe
Now we see the point.  If $\delta_i > D$, this term becomes less and less
important at low energies, and so is termed {\it irrelevant}.  Similarly, if
$\delta_i < D$, the operator is more important at lower energies and is termed
{\it relevant}.  An operator with $\delta_i = D$ is equally important at all
scales and is {\it marginal}.  This is summarized in the table below,
along with the standard terminology from renormalization theory.
\begin{center}
\begin{tabular}{ccccc}
$\delta_i$ & size as $E \to 0$ &&& \\ \hline
$< D$ & grows & relevant && superrenormalizable \\[2pt]
$= D$ & constant & marginal && strictly renormalizable \\[2pt]
$> D$ & falls & irrelevant && nonrenormalizable
\end{tabular}
\end{center}
In most cases there is only a finite number of relevant and marginal terms,
so the low energy physics depends only on a finite number of parameters.
For example, one sees from the dimension~\ee{scalopdim} that this is true
of scalar field theory in $D \geq 3$.

Why do we emphasize the
free action in determining the dimensions of the fields?  It is because we are
assuming that the theory is weakly coupled, so that the free action determines
the sizes of typical fluctuations, or matrix elements, of the fields; later we
will discuss corrections to this.  It is necessary here that the coefficient of
the dominant term  in the action is made dimensionless, as in the
example~\ee{scakin}, by rescaling of the fields.  This is used in the
estimate~\ee{opest}, where it is implicitly assumed that the only dimensionful
quantity is the energy scale.

For more general applications, it is useful to state things in a way that does
not depend on dimensional analysis.  Again assume that the kinetic
term~\ee{scakin} is dominant.  Imagine scaling all energies and momenta by a
factor $s < 1$, so lengths and times scale by $1/s$.  The volume element and
derivatives in the kinetic term scale as 
$s^{2-D}$, so the fluctuations of $\phi$
scale as $s^{- 1 + D/2}$ and the $i$'th interaction 
then scales as $s^{\delta_i -
D}$, thus reproducing the earlier conclusion about relevance and irrelevance.
In some contexts there are two `kinetic terms.'  For example, there can be
both first derivative Chern-Simons and second derivative Maxwell terms present
in $2+1$ dimensional gauge theory, but at any given momentum one will dominate
the other and determine the scaling.  Similarly in statistical mechanics of
membranes, there can be both second derivative tension and fourth derivative
rigidity terms.

There are many comments and elaborations to make, but let us first
list some classic examples:\\[6pt]
\begin{tabular}{lcl}
High Energy Theory & $E_0$ & Low Energy Theory \\ \hline
1. Weinberg-Salam & $M_W \sim$ 80 GeV & Fermi weak interaction theory \\[2pt]
2. grand unified theory & $M_{\slm GUT} \sim 10^{16}$ GeV & $SU(3) \times
SU(2) \times U(1)$ \\
3. QCD & $M_\rho \sim .8$ GeV & current algebra \\[2pt]
4. lattice field theory & -- & continuum field theory \\[2pt]
5. string theory & $M_{\slm string} \sim 10^{18}$ GeV & field theory of gravity
and matter
\end{tabular}\\[6pt]

In the first two examples, both the high and low energy theories are
perturbative field theories.  Notice in the first example that there
is {\it no} relevant or marginal weak interaction in the low energy theory. The
largest irrelevant term is dimension~6, suppressed by two powers of $E_0$, but
of course it still has observable effects: `irrelevant' is not to be taken in
precisely its colloquial sense.  We also should
note that the simple frequency separation~\ee{momsep} is usually an impractical
way to calculate.  Life would be much
easier if we could use dimensional regularization,
rather than a cutoff, in the low energy theory.  
Now, dimensional regularization
is not well-suited conceptually to our
discussion, but once we have decided that
the low energy field theory exists we are free to use any regulator we want.
The point is that we know from renormalization theory that the effect of
changing from a cutoff regulator to some other can be absorbed into the
parameters $g_i$ in the Lagrangian. The values of the $g_i$ appropriate for any
given regulator are determined by a matching calculation: calculate some
amplitude in the full theory, and in the 
low energy theory, and fix the $g_i$ so
that the answers are the same.

The third example illustrates several further points.  First, the right fields
to use in the low energy theory are not the same as those in the high energy
theory.  If we simply carried out the frequency splitting in terms of the gluon
and quark fields, the resulting low energy theory would be too complicated to
be useful; instead we need the composite fields $\bar q q$.  When the theory
is strongly coupled as here, we need to find the right fields to give a simple
description.  The only low energy degrees of freedom surviving below the QCD
scale are those guaranteed by Goldstone's 
theorem, so the appropriate low energy
field is the local alignment of the $SU(3) \times SU(3) \to SU(3)$ breaking.
Second, because the theory is strongly coupled, we cannot calculate the low
energy theory directly: we have to determine the $g_i$ empirically, or
by some model or Monte Carlo calculation.  Third, it is not necessary
to have an enormous ratio of scales for the effective Lagrangian to be useful.
In $s$-quark current algebra, the ratio is $M_K / M_{K^*} \sim 0.6$.

In the fourth example the short distance field theory is not even
local, but the effective theory at long distance is a local continuum theory.
In the final example, too, spacetime breaks down in some ill-understood
way at short distance.  Also, it is not clear that a path integral
representation~\ee{pathint}---that is, a string field theory---is the best
description of the short distance theory, but all indications are that the long
distance theory is an ordinary quantum field theory.

Now for some ideology.  Presumably no field theory we have ever encountered,
and perhaps no field theory of any type, is complete up to arbitrarily high
energies.  They are all effective theories, valid up to some cutoff.  If there
is a cutoff, does this mean that renormalization is irrelevant (in the
colloquial sense)?  Not at all; the results are just as important but the
meaning is different.  Rather than the `cancellation of infinities' that has
always seemed so artificial and is still taught in most textbooks, the meaning
is much more physical.  It was stated above, but is repeated here for
emphasis:\\[4pt]
{\it The low energy physics depends on the 
short distance theory only through the
relevant and marginal couplings, and possibly through some leading irrelevant 
couplings if one measures small enough effects.}\\[4pt]
\indent
The power counting above makes this seem obvious, but there are subtleties,
due to divergences in the low energy field theory.  An irrelevant operator
comes with a negative power of the cutoff, but if embedded in a divergent
graph this factor could be offset.  It is easy to extend the usual
renormalization power counting to show that any such divergence with an overall
non-negative power of the cutoff has the form of a relevant or marginal
operator, and so can be absorbed into those couplings.  Going further, we have
to justify the naive power counting.  In the usual approaches, this is a
combinatoric challenge, involving skeleton expansions, overlapping divergences,
and so forth.  One might suppose that any result that seems so obvious and
dimensional should have a simple and general proof.  In fact, that is the
case.  The result becomes obvious if one thinks about doing the path integral
one frequency slice at a time: first over $\Lambda > \omega > \Lambda -
d\Lambda$, then over $\Lambda - d\Lambda > \omega > \Lambda - 2d\Lambda$,
and so forth.  This generates the Wilson equation, a differential equation for
the action,
\eq
\frac{\del S_\Lambda}{\del \Lambda} = {\cal F}(S_\Lambda), \e{wilson}
\eqe
where $\cal F$ is some functional.  The 
derivation, and the explicit form of $\cal F$,
are rather simple and can be found in the references.  The Wilson equation is a
flow in an infinite dimensional space.  Linearizing around a solution,
irrelevant operators
correspond to negative eigenvalues, 
directions in which the flow is converging and erasing
information about the initial conditions.  Now, if we linearize
around {\it zero coupling} the eigenvalues 
are given precisely by power counting
as in eq.~\ee{scalopdim}, $D - \delta_i$.  The right-hand side 
of the Wilson equation is a smooth function of the
couplings.  There is no place for a singularity to come from, because we are
doing a path integral only over a frequency range $d\Lambda$, with both an IR
and a UV cutoff.  So eigenvalues which are negative in the free theory remain
negative for sufficiently small nonzero
couplings---QED.  I must confess that in attempting to turn this argument into
some precise inequalities (Polchinski, 1984), I made things look much less
transparant than they really are, but the argument above is really all there is
to it---no skeletons, no overlapping divergences.

This discussion does bring up an important 
point, that interactions modify the naive
scaling~\ee{opest} found from the free action.   
At sufficiently strong coupling,
a finite number of interactions will 
in some cases switch between relevant, marginal, and irrelevant,
so that the theory is still determined by 
a finite number of parameters but the number
differs from the naive perturbative count.  The 
Thirring model is a simple example of
this.  Another is `walking technicolor,' in which 
it is supposed that an irrelevant couplng
is enhanced to nearly marginal by interactions. 
Yet another is the infrared fixed point in $D=3$ scalar field theory,
responsible for the critical behavior in many systems.  At weak coupling, the
$\phi^4$ interaction is relevant in $D = 3$, but at sufficiently strong 
coupling this is offset by the quantum effects and there is a zero of
the beta function.

Of course, the corrections to naive scaling 
are most important for marginal couplings,
since an arbitrarily small correction 
will make these relevant or irrelevant.  The
effective value of a single marginal coupling $g$ will behave something like
\eq
E\del_E g = b g^2 + O(g^3).  \e{beta}
\eqe
If $b > 0$, the coupling decreases with decreasing 
energy and is marginally irrelevant;
if $b < 0$, the coupling grows and is 
marginally relevant; if $b$ and all higher terms
vanish, the coupling is truly marginal. 

Let us emphasize somewhat further the marginally relevant case.  Integrating
eq.~\ee{beta} gives
\eq
g(E) = \frac{g(\Lambda)}{1 + b g(\Lambda) \ln(\Lambda / E)},
\eqe
For $b < 0$ the coupling grows strong at $E 
\sim \Lambda e^{1/bg(\Lambda)}$.  What
happens next depends on the details of the problem.  In QCD, what happens is
confinement and chiral symmetry breaking.  
In technicolor models, what happens is
$SU(2) \times U(1)$ breaking.  In models 
of dynamical supersymmetry breaking what
happens is spontaneous breaking of supersymmetry 
and subsequently of $SU(2) \times
U(1)$.   This general pattern, a marginal
coupling growing strong and then something
interesting happening, is a simple means 
of generating interesting dynamics and large
ratios of scales.  We will see several further examples in the second lecture.

Now for some more ideology.  In contrast 
to textbook renormalization theory, where
nonrenormalizable terms are strictly forbidden, 
we always expect nonrenormalizable
terms to appear at some level.  They are 
harmless because the effective theory has a
cutoff, an in fact they tell us where the cutoff must appear.  If we observe a
nonrenormalizable coupling $g$ with units 
$E^{D - \delta}$, $\delta > D$, the effective
dimensionless coupling $g E^{\delta - D}$ 
grows with increasing energy.  Presuming that
the effective theory breaks down when the coupling greatly exceeds 1, we
have $\Lambda < O( g^{1/(\delta - D) } )$.

Nonrenormalizable terms are not a 
problem, but there is a new sort of problem:
superrenormalizable terms!  To see why these are bad, consider the operator
$\phi^2$ in scalar field theory, which has 
$D - \delta = 2$.  This appears in the action
with coefficient $\lambda_{\phi^2} \Lambda^2$.
The path integral~\ee{highint} which produces the effective action in general
generates all terms allowed by symmetry, with dimensionless coefficients of
order~1.\footnote{There are some specific exceptions---certain topological
terms and certain supersymmetric terms.}  To produce a much smaller coefficient
requires an unnatural fine-tuning of the parameters in the original theory.
Without fine-tuning, the dimensionless $\lambda_{\phi^2}$ will
be of order~1.  But this is a
contradiction: it is a mass term of 
order the cutoff, so that $\phi$ does not appear in
the low energy theory at all!  So we have 
a new condition: effective field theories must
be {\it natural}, meaning that all possible masses
must be forbidden by symmetries.
The
textbook criterion for a consistent theory, renormalizability, 
is automatically satisfied
in an effective theory for dimensional 
reasons, but the new and stringent criterion of
naturalness appears.  A natural effective theory 
may have gauge interactions, because 
a mass is forbidden by gauge invariance.  
It may have fermions, if their masses are
forbidden by chiral symmetry, and scalars, if their masses are forbidden by
supersymmetry or Goldstone's theorem.

Masses are obviously bad, but it is less obvious
whether superrenormalizable {\it interactions} are also
bad.  At scales below the cutoff, a superrenormalizable coupling will have a 
dimensionless strength much greater than unity.  One would therefore expect
complicated dynamics, likely with the formation of bound states and condensates.
The low energy theory will then look very different
from what appears in the Lagrangian (as is the case, for example, when
the QCD coupling gets strong), and one should find a new effective theory which
more accurately describes the low energy degrees of freedom.  There is at least
one sort of exception to this reasoning.  In $D = 3$ scalar field theory, with
the mass tuned to zero, the interaction is relevant at weak coupling,
but as discussed above the infrared behavior is governed by a fixed point.
This is not a free theory, but the degrees of freedom at the fixed point are
still those of a scalar field.

Is the Standard Model natural?  No.  A 
mass-squared for the Higgs scalar doublet
is not forbidden by any symmetry of the theory.  By the reasoning above, it is
a contradiction to suppose that the Standard Model is valid to scales far above
the weak scale.  Rather, we 
must soon find some new theory, either without scalar
fields (technicolor) or with a symmetry that makes it natural (supersymmetry).
This is the eight billion 
dollar argument---of course we expect something beyond
the Standard Model at some energy, but the naturalness argument is the only one
which says that new physics will appear at energies accessible to accelerators.

There is one other relevant operator possible in 
the Standard Model, the operator
1, which has dimension zero.  
When the coupling to gravity is taken into account,
this is a cosmological constant.  Obviously 1 is rather symmetric and hard to
forbid.  Supersymmetry can do it, but to suppress the cosmological constant
to the necessary degree this would have to be unbroken down to around $10^{-3}$
eV, which is obviously not the case.  This is a hard problem.

{\bf Q:}  Doesn't the infinite 
sum in the effective action~\ee{effact} mean that
there is no predictive power?

{\bf A:}  No, rather effective field theory gives an accurate statement of how
much predictive power one really 
has.  For example, if I tell you that physics at
the weak scale is described by the 
Standard Model, you can't calculate the proton
lifetime.  But if you know that the Standard Model is a valid effective field
theory up to $10^{16}$ GeV (ignoring, for the sake of illustration, the
naturalness problem), you can say that the proton decay rate is zero to an
accuracy of order $(1 {\ \rm GeV} )^5 / (10^{16} {\ \rm GeV})^4 $.  
In a similar
way, all predictions are accurate to a known power of the inverse cutoff.

{\bf Q:}  Doesn't all this mean that 
quantum field theory, for all its successes,
is an approximation that may have little to do with the underlying theory?
And isn't renormalization a bad thing, since it implies that we can only
probe the high energy theory through a small number of parameters?

{\bf A:}  Nobody ever promised you a rose garden.
 
\section*{Lecture 2: The Landau Theory of Fermi Liquids}

Now, as an illustration, we will derive the effective field theory of the low
energy excitations in a conductor.  As far as I know, this subject is not
presented in this way anywhere in the literature.  However, it is clear that
the essential idea is entirely familiar to those in the field.  It
is implicit in the writings of Anderson, and recently has been made more
explicit by Benfatto and Gallavotti, and Shankar.

Before starting, let me describe one chain of thought which led me to this.
Introductions to superconductivity often point out 
the following remarkable fact.
In ordinary metallic superconductors, the size of a Cooper pair may be as 
large as $10^{4}$ \AA.
The orbital thus overlaps $10^{10}$ or more other electrons, with the
characteristic electronic interaction energies being as large 
as an electron volt or more.
The BCS theory neglects all but the binding interaction between the
paired electrons, which is of order $10^{-3}$ eV. 
Yet this leads to results which are not only qualitatively correct but
also quantitatively: calculations in BCS theory are supposed to be accurate to
order $(m/M)^{1/2}$, where $m$ is the electron mass and $M$ the nuclear mass.  
Further, while the BCS theory is for weak electron-phonon coupling, 
it has a strong-coupling generalization, Eliashberg theory, which for arbitrary
coupling remains valid to accuracy $(m/M)^{1/2}$.  This is again remarkable,
because solvable field theories in $3+1$ dimensions are few and far between.

The BCS and Eliashberg theories are derived within the Landau theory of Fermi
liquids, which treats a conductor as a gas of nearly free electrons.  The
justification for this appeals to the notion of `quasiparticles,' which are
dressed electrons, the neglected interactions being absorbed in the
dressing.  The term quasiparticle is not 
in common use in particle theory, nor is
the notion that a strongly interacting theory can be turned into a nearly free
one just by such dressing.  What we will see is that in fact this theory is a
beautiful example of an effective field theory.  The neglected interactions can
be regarded as having been {\it integrated out}, in the usual effective field
theory sense.  This is possible because of the special kinematics of the Fermi
surface.  Further, the resulting theory is 
solvable because there are {\it almost
no} relevant or marginal interactions, in a sense that will be made clear.

We begin by identifying the characteristic scale.  The electronic properties of
solids are determined by $e$, $\hbar$ and $m$, out of which we can construct
the energy $e^4 m / \hbar^2 = 27$ 
eV.   Typical electronic energies, such as the
width of the conduction electron band, are actually slightly smaller than
this, say $E_0 \sim 1$ to 10 eV.  The other possible constants, $M$
and $c$, are so much greater than the electron mass and velocity that we can
treat them as infinite.  Later we will introduce $1/M$ effects
(lattice vibrations).  We will see that the fact that solids are near the
$M = \infty$ limit is a great simplification.
Spin-orbit coupling and other $1/c$ effects also are important in some
situations, but not for our discussion.

In a conductor, we can 
excite a current with an arbitrarily weak electric field,
so the spectrum of charged excitations evidently goes down to zero energy.
This is the hierarchy of scales that 
makes effective field theory a useful tool;
we want the effective theory describing the excitations with $E << E_0$.  Now,
at $E_0$ there are electrons with strong Coulomb interactions.  We
are not going to try to solve this theory.  
Rather, we are in a situation similar to
the current algebra example, where we will write down the most general
effective theory with given fields and symmetries. 
At this point we need to make a guess---what are the light charged fields?
Let us suppose that they are spin-$\frac{1}{2}$ fermions, like the underlying
electrons.\footnote{I will therefore call them electrons.}  It must be
emphasized that this is only a guess.  It can be justified in the artificial
limit of a very dilute or weakly interacting system, but with strong
interactions it is possible that something very different might emerge.  All we
can do here is to check the guess for consistency (naturalness), and compare it
with experiment.

Begin by examining the free action
\eq
\int dt\, d^3 \bp \, \Bigl\{  i 
\psi_\sigma^\dagger(\bp) \del_t \psi_\sigma(\bp)
- (\varepsilon(\bp) - \varepsilon_{\slm F})
\psi_\sigma^\dagger(\bp)\psi_\sigma(\bp) \Bigr\}. \e{act1}
\eqe
Here $\sigma$ is a spin index and $\varepsilon_{\slm F}$ is the Fermi energy.
The single-electron energy $\varepsilon(\bp)$ would be $p^2 / 2m$ for
a free electron, but in the spirit of 
writing down the most general possible action
we make no assumption about its form.\footnote
{A possible $\bp$-dependent
coefficient in the time-derivative term has been 
absorbed into the normalization
of $\psi_\sigma(\bp)$.}
The ground state of this theory is the Fermi sea, with
all states $\varepsilon(\bp) < \varepsilon_{\slm F}$ filled and all states
$\varepsilon(\bp) > \varepsilon_{\slm F}$ 
empty.  The Fermi surface is defined by
$\varepsilon(\bp)  = \varepsilon_{\slm F}$.  Low 
lying excitations are obtained by adding
an electron just above the Fermi 
\begin{figure}
\begin{center}
\leavevmode
\epsfbox{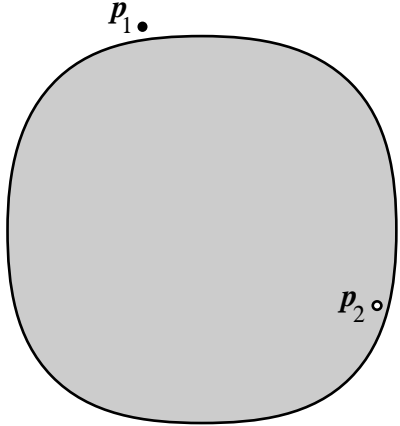}
\end{center}
\caption[]{Fermi sea (shaded) with two low-lying excitations, an electron at
$\bp_1$ and a hole at $\bp_2$.}
\end{figure}
surface, or removing one (producing a hole) just
below, as shown in figure~1.

Now we need to ask how the fields behave as we scale all energies by a factor
$s < 1$.  In the relativistic case, the 
momentum scaled with the energy, but here
things are very different.  As figure~1 
makes clear, as the energy scales to zero
we must scale the momenta {\it toward the Fermi surface}.  To do this, write
the electron momentum as
\eq
\bp = \bk + \bl,
\eqe
where \bk is vector on the Fermi surface and \bl is a vector orthogonal to the
Fermi surface.  Then when $E \to sE$, the momenta scale $\bk \to \bk$ and 
 $\bl \to s\bl$.  Expand the single particle energy
\eq
\varepsilon(\bp) - \varepsilon_{\slm F} = l v_{\slm F}(\bk) + O(l^2),
\eqe
where the Fermi velocity ${\bf v}_{\slm F} = \del_\bp \varepsilon$.
Scaling
\eq
dt \to s^{-1} dt, \quad d\bk \to d\bk, \quad d\bl \to s d\bl, \quad \del_t 
\to s \del_t, \quad l \to sl,
\eqe
each term in the action
 \eq
\int dt\, d^2 \bk \, d\bl \, \Bigl\{ i \psi_\sigma^\dagger(\bp) \del_t
\psi_\sigma(\bp) - l v_{\slm F}(\bk)
\psi_\sigma^\dagger(\bp)\psi_\sigma(\bp) \Bigr\} \e{act2}
\eqe
scales as $s^{1}$ times the scaling of $\psi^\dagger
\psi$.  The fluctuations of $\psi$ thus scale as $s^{-1/2}$. 

Now we play the effective field theory game, writing down {\it all} 
terms allowed
by symmetry and seeing how they scale.  If we find a relevant term we lose:
the theory is unnatural.  The symmetries are
\begin{itemize}
\item[1.] Electron number.
\item[2.] The discrete lattice symmetries.  Actually, in the action~\ee{act1},
we have treated translation invariance as a continuous symmetry, so that
momentum is exactly conserved.  Because the electrons are moving in a periodic
potential, they can exchange discrete amounts of momentum with the lattice. 
Including these terms, the free action can be rediagonalized, with the result
that the integral over momentum becomes a sum over bands and an integral over a
fundamental region (Brillouin zone) for each band.  This does not affect the
analysis in any essential way, so for simplicity we will treat momentum as
exactly conserved.  In addition, the action is constrained by any discrete
point symmetries of the crystal.
\item[3.] Spin $SU(2)$.  In the $c \to \infty$ limit, physics is invariant
under independent rotations of space and spin, so spin $SU(2)$ acts as an
internal symmetry.
\end{itemize}

Starting with terms quadratic in the fields, we have first
\eq
\int dt\, d^2 \bk \, d\bl \, \mu(\bk) \psi_\sigma^\dagger(\bp) 
\psi_\sigma(\bp).
\eqe
Combining the scaling of the various factors, this goes as $s^{-1 + 1 -2/2} =
s^{-1}$.  This resembles a mass term, and it is relevant.  Notice, though, that
it can be absorbed into the definition of $\varepsilon(\bp)$.  We should expand
around the Fermi surface appropriate to the full $\varepsilon(\bp)$.  Thus, the
{\it existence} of a Fermi surface is natural, but it is unnatural to assume it
to have any very precise shape beyond the constraints of symmetry.  Adding one
time derivative or one factor of $l$ makes the operator marginal, scaling as
$s^0$; these are the terms already included in the action~\ee{act2}.
Adding additional time derivatives or factors of $l$ makes an irrelevant
operator.

Turning to quartic interactions, the first is
\eqa
&& \int dt\, d^2 \bk_1 \, d\bl_1 \, 
d^2 \bk_2 \, d\bl_2 \, d^2 \bk_3 \, d\bl_3 \,
d^2 \bk_4 \, d\bl_4 \, V(\bk_1,\bk_2,\bk_3,\bk_4) \e{fourfermi}\\[2pt]
&&\qquad\qquad
\psi_\sigma^\dagger(\bp_1) \psi_\sigma(\bp_3) \psi_{\sigma'}^\dagger(\bp_2)
\psi_{\sigma'} (\bp_4) \delta^3 (\bp_1 + \bp_2 - \bp_3 - \bp_4 ).
\nonumber
\eqae
This scales as $s^{-1 + 4 - 4/2} = s$, times the scaling of the delta-function.
Let us first be glib, and argue that 
\eqa
\delta^3 (\bp_1 + \bp_2 - \bp_3 - \bp_4 ) &=& \delta^3 (\bk_1 + \bk_2 - \bk_3
- \bk_4 + \bl_1 + \bl_2 - \bl_3 - \bl_4 ) \nonumber\\[3pt]
&\sim& \delta^3 (\bk_1 + \bk_2 -
\bk_3 - \bk_4 ). \e{deltanaive}
\eqae
That is, we ignore the $\bl$ compared to the $\bk$, since the former are
scaling to zero.  The argument of the delta-function then does not depend on
$s$, so the delta-function goes as $s^0$ and the overall scaling is $s^1$.
Pending a more careful treatment later, 
the operator~\ee{fourfermi} is irrelevant.  It is then
easy to see that any further interactions are even more irrelevant.

To summarize, with our assumption about the nature of the charge carriers
the most general effective theory has only irrelevant interactions, becoming
more and more free as $E \to 0$.  The assumption of a nearly
free electron gas is thus internally consistent, and in fact is a good
description of most conductors.  It should be emphasized that this is just a
reformulation of a simple and standard solid state argument, to the effect that
the kinematics of the Fermi surface plus the Pauli exclusion principle
exclude all but a fraction $E / E_{0}$ of possible final states in any
scattering process.

There are two complications to discuss before the analysis is complete.
The first is phonons.   Because a crystal spontaneously breaks spacetime
symmetries, the low energy theory must include the correponding Goldstone
excitations.  We therefore introduce a phonon field $\bD(\br)$,
which is equal to $M^{1/2}$ times the local displacement of the ions from
their equilibrium positions.\footnote{
Notice that a crystal actually breaks {\it nine} spacetime symmetries: three
translational, three rotational and three Galilean.  For internal symmetries,
Goldstone's theorem gives a one-to-one correspondence between broken symmetries
and Goldstone bosons.  This 
need not be true for spacetime symmetries, and three
Goldstone fields are sufficient to saturate all the broken symmetry Ward
identities. } The free action for this field is
\eq
\frac{1}{2} \int dt\, d^3 \bq\, \Bigl\{ \del_t D_i (\bq) \del_t D_i (-\bq)
- M^{-1} \Delta_{ij}(\bq) D_i (\bq) D_j (-\bq) \Bigr\}. \e{freepho}
\eqe
In the time derivative term, the $M$ in the kinetic energy of the ions
cancels against the factors of $M^{-1/2}$ from the definition of $\bD$;
again we have made a finite rescaling to eliminate a $\bq$-dependent
coefficient.  The restoring force, on the other hand, is to first approximation
independent of $M$ and so the $M^{-1}$ comes from the definition of $\bD$.

Now examine the scaling of the phonon field.  We determine the scaling from the
kinetic term; except at very low frequenies (to be discussed) this
term dominates because of the $M^{-1}$ in the restoring force.  How does
$\bf q$ scale?  Figure 2 shows a typical phonon-electron interaction, such as
is responsible for BCS superconductivity.  
\begin{figure}
\begin{center}
\leavevmode
\epsfbox{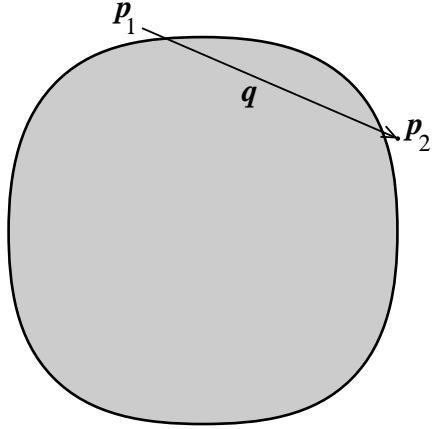}
\end{center}
\caption[]{An electron of momentum $\bp_1$ absorbs a phonon of large momentum
$\bq$ but remains near the Fermi surface.}
\end{figure}
As the electron momenta scale
towards the Fermi surface, $\bq$ approaches a nonzero limit, so $\bq \propto
s^0$.  The integration and time derivatives in the kinetic term~\ee{freepho}
then scale as $s^1$, so the phonon field scales as $s^{-1/2}$.

The restoring force is relevant, scaling as $s^{-2}$, so in spite of its small
coefficient it will dominate at energies below
\eq
E_1 = (m/M)^{1/2} E_0.
\eqe
This is the Debye energy, the characteristic energy scale of the phonons.
The restoring force is like a mass term, making the phonons decouple below
$E_1$.  Of course, Goldstone's theorem guarantees that the eigenvalues of
$\Delta_{ij}(\bq)$ vanish as $\bq \to 0$, so there are still some phonons
present at arbitrarily low energy.  But their effects are doubly suppressed,
by the phonon phase space and because, as Goldstone bosons, their interactions
are proportional to $\bq$.  The long-wavelength phonons can therefore be
neglected for most purposes.

Now consider interactions, starting with
\eqa
&& \int dt\, d^3 \bq\, d^2 \bk_1\, d\bl_1 \, d^2 \bk_2 \, d\bl_2 \,
M^{-1/2} g_i (\bq, \bk_1, \bk_2) \e{dee}\\[2pt]
&&\qquad\qquad\qquad\qquad 
D_i (\bq) \psi_\sigma^\dagger (\bp_1) \psi_\sigma(\bp_2)
\delta^3 (\bp_1 - \bp_2 - \bq), \nonumber
\eqae
where an electron emits or absorbs a phonon.
The electron-ion force is to first approximation independent of $M$, so the
explicit $M^{-1/2}$ is from the definition of $\bD$.  This scales as
$s^{-1 + 2 - 3/2} = s^{-1/2}$ if we treat 
the delta-function as before, so it is
relevant.  When the phonons decouple at $E_1$, the 
coupling has grown by $(E_1 /
E_0)^{-1/2} = (m / M)^{-1/4}$. However, combined with the small dimensionless
coefficient $(m/M)^{1/2}$ of the interaction~\ee{dee}, this leaves an overall
suppression of $(m/M)^{1/4}$.

There are two ways to proceed to lower energies.  The first is simply to note
that the restoring force dominates the 
kinetic term below $E_1$, and so should be
used to determine the scaling.  Then $\bD$ scales as $s^{+1/2}$ and so does the
interaction~\ee{dee}; it is irrelevant below $E_1$.  Alternately, we can
integrate the phonon out.  The leading interaction induced in this way
is again the four-Fermi term~\ee{fourfermi}, which by the earlier analysis is
irrelevant.
Further interactions are even more suppressed, so the inclusion of phonons has
not changed the free electron picture: we find an electron-phonon interaction
which reaches a maximum of order $(m/M)^{1/4}$ at the Debye energy and then
falls.

If this were the whole story, it would be rather boring.  It is difficult
to see how we can ever get an interesting collective effect like
superconductivity in the low energy theory if all interactions are getting 
weaker and weaker.
However, there is an important subtlety in the kinematics, so that our
treatment~\ee{deltanaive} of the delta-function is not always valid.
This simplest way to see this is pictorial (figure 3).  
Consider a process where
electrons of momenta $\bp_{1,2}$ scatter into momenta $\bp_{3,4}$.  Expand
\eq
\bp_3 = \bp_1 + \delta\bk_3 + \delta\bl_3, \qquad
\bp_4 = \bp_2 + \delta\bk_4 + \delta\bl_4.
\eqe
The momentum delta-function in $d_{\slm s}$ space dimensions is then
\eq
\delta^{ d_{\slm s} } (\delta\bk_3 + \delta\bk_4 + \delta\bl_3 + \delta\bl_4).
\eqe
Now, for generic momenta, shown in figure~3a, $\delta\bk_3$ and $\delta\bk_4$
are linearly independent and our neglect of $\delta\bl_3$ and $\delta\bl_4$
is justified.  
An electron of momentum $\bp_1$ absorbs a phonon of large momentum
$\bq$ but remains near the Fermi surface.
\begin{figure}
\begin{center}
\leavevmode
\epsfbox{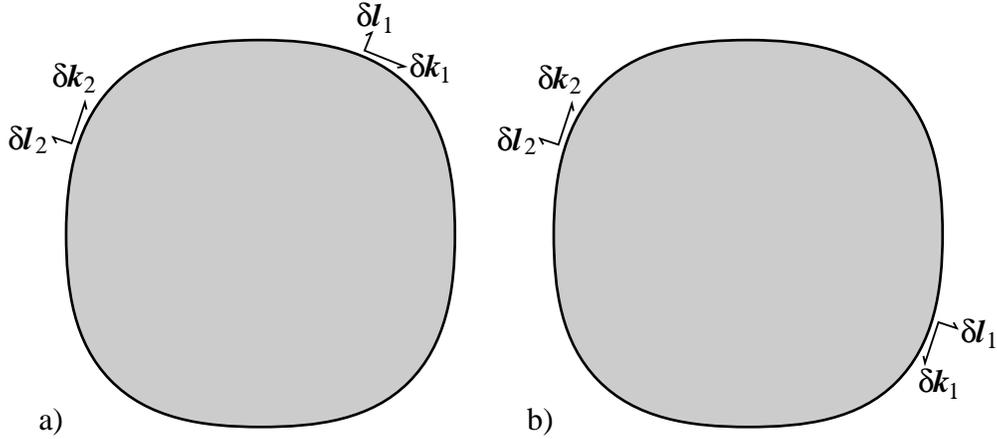}
\end{center}
\caption[]{a) For two generic points near a two-dimensional Fermi surface, the
tangents~$\delta\bk_i$ are linearly independent. b) For diametrically opposite
points on a parity-symmetric Fermi surface, the tangents are parallel.}
\end{figure}
Incidentally, while the picture is two-dimensional, it is easy
to see that this argument applies equally for all $d_{\slm
s} \geq 2$: the possible variations $\delta\bk_3$, $\delta\bk_4$ span the full
$d_{\slm s}$-space.  However, if $\bp_1 = -\bp_2$, so that 
the total momentum is
zero, then $\delta^{ d_{\slm s} } (\delta\bk_3 + \delta\bk_4)$ is {\it
degenerate}, since one component of the argument vanishes automatically.
In this case, one component of the delta-function {\it does} constrain the
$\bl$, and so scales inversely to $\bl$, as $s^{-1}$.  The four-Fermi
interaction then scales as $s^0$; it is {\it marginal}.\footnote{Notice that we
implicitly assume a discrete symmetry, namely parity invariance of the Fermi
surface.  Incidentally, one must be a bit careful.  One would seem to find the
same enhancement for $\bp_1 = +\bp_2$.  In that case, however, the
delta-function is degenerate only at one point on the Fermi surface, so that 
second order terms in $\delta\bk$ are nonzero and the enhancement is only by
$s^{-1/2}$.}

The rule which emerges is that in any process, if the external momenta are such
that the total momentum $\bP$ of two the lines in a four-Fermi vertex is
constrained to be zero, that vertex is marginal.\footnote
{If $\bP$ is not exactly zero, the interaction is marginal for $E > v_{\slm F}
P$ and irrelevant for $E < v_{\slm F} P$. }
All other fermionic interactions
remain irrelevant.  To treat the phonons, the most efficient approach seems to
be to consider the effective four-Fermi interaction induced by phonon exchange,
and then the same rule applies.  We apportion the enhancement as a factor
of $s^{-1/2}$ in each vertex, so the phonon-electron interaction scales as
$s^{-1}$ above $E_1$.  At $E_1$, it is then of order $(m/M)^{1/2 - 1/2}$,
unsuppressed in the $M \to \infty$ limit.

The existence of a marginal interaction only at special points in momentum
space leaves the free-Fermi picture largely intact, but there are important
changes.  Consider the matrix element of some current between electrons
of momenta $\bp$ and $\bp'$.  The tree level graph is shown in figure~4a,
and a one-loop correction in figure~4b.  
\begin{figure}
\begin{center}
\leavevmode
\epsfbox{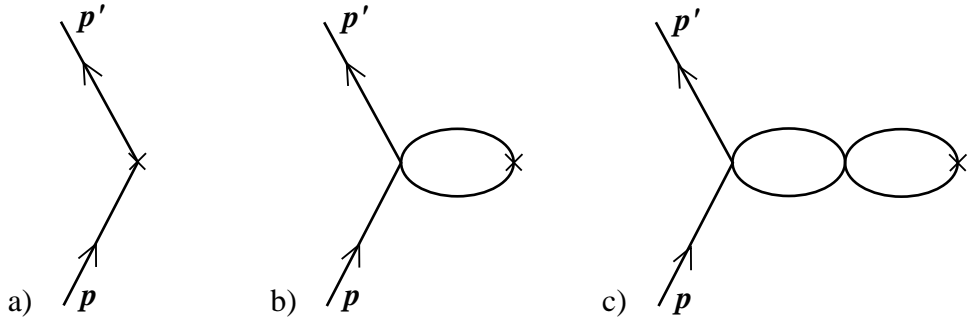}
\end{center}
\caption[]{a) Tree-level matrix element of current. b) One-loop correction
which is marginal at $\bp = \bp'$. c) Two-loop correction which is marginal
at $\bp = \bp'$.}
\end{figure}
If $\bp$ and $\bp'$ are both near 
the Fermi surface but their difference is not small, then the interaction in
figure~4b is irrelevant and the loop correction small.  In an expectation
value, however, where $\bp = \bp'$, the interaction is marginal
and the loop graph is unsuppressed near the Fermi surface.  Similarly,
both interactions in the two-loop graph of figure~4c are marginal, and so on
with any number of bubbles in the chain.  The graphs with no irrelevant
interactions thus form a geometric series.  This is the Landau theory of
Fermi liquids: expectation values of currents are modified from their
free-field values by the interaction.

The same consideration applies to the electron-phonon vertex.  Imagine coupling
in a phonon where the current appears in figure~4.  As in the discussion of
figure~2, the typical phonon momentum $\bq$ is not small, so the interactions
are irrelevant and only the tree level graph~4a contributes.  This is Migdal's
theorem.

Another way to think about the situation is that the interaction is always
irrelevant and decreases with $E$, but for special kinematics 
an infrared divergence comes in to precisely offset this.
We should emphasize the dependence on dimension.  The analysis thus far
is valid for all $d_{\slm s} \geq 2$.  For one spatial dimension, however,
there is no $\bk$, only $\bl$.  The delta-function then {always} scales as
$s^{-1}$, and the four-Fermi interaction is always marginal.  In this case,
there is no simplification of the theory---no irrelevant graphs and no Migdal's
theorem---and there is more possibility of interesting dynamics.

As was discussed in lecture~1, when we have an interaction which is
classically marginal it is important to look at the quantum corrections.
Figure~5 shows the four-Fermi vertex and the one-loop correction.  
\begin{figure}
\begin{center}
\leavevmode
\epsfbox{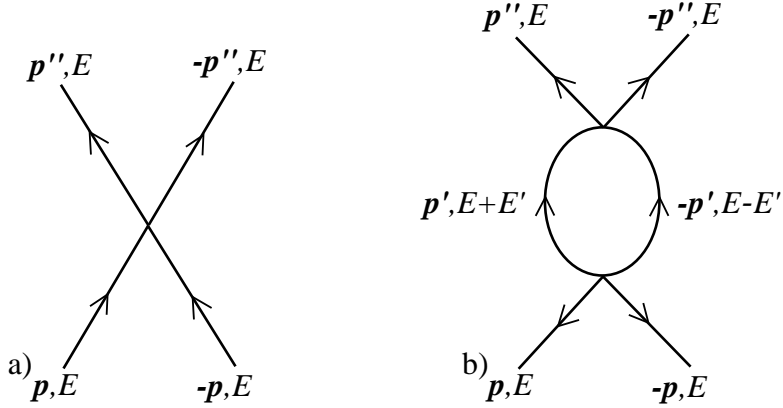}
\end{center}
\caption[]{Scattering of electrons $(\bp, E)$ and $(-\bp, E)$
to $(\bp'', E)$ and $(-\bp'', E)$. a) Tree level. b) One loop.}
\end{figure}
The Feynman
rules are easily worked out.  It is convenient to focus on the case that
$V(\bk_1,\bk_2,\bk_3,\bk_4)$ is a constant, which is an approximation often
made in practice.  Then the one loop four-Fermi amplitude of figure~5b is 
\eq
V^2 \int \frac{dE'\, d^2\bk'\, dl'}{(2\pi)^4} \frac{1}
{ \Bigl[ (1+i\epsilon)(E + E') - v_{\slm F}(\bk') l' \Bigr]
  \Bigl[ (1+i\epsilon)(E - E') - v_{\slm F}(\bk') l' \Bigr] }.
\eqe
We are only interested in the large logarithm, and so do
not need to know the details of how the upper cutoff at $E_0$ is implemented.
Evaluating the integral to this accuracy gives
\eq
V(E) = V - V^2 N \Bigl\{ \ln(E_0 / E) + O(1) \Bigr\} + O(V^3),
\eqe
where
\eq
N = \int \frac{d^2\bk'}{(2\pi)^3} \frac{1}{ v_{\slm F}(\bk') }
\eqe
is the density of states at the Fermi energy.
Inserting into the renormalization group~\ee{beta} for $V$, one determines
$b = N$:
\eq
E \del_E V(E) = N V^2(E) + O(V^3),
\eqe
with solution
\eq
V(E) = \frac{V}{1 + NV \ln (E_0 / E)}. \e{running}
\eqe
A repulsive interaction ($V > 0$) thus grows weaker at low energy, while an
attractive interaction ($V < 0$) grows stronger.

Now we are in a position to learn something interesting.  We start at $E_0$
with a four-Fermi coupling $V_{\slm C}$ and the phonon-electron coupling $g$,
where we again for simplicity ignore the $\bk$ dependences.  The subscript C is
for Coulomb, since this is some sort of screened Coulomb interaction.  Defining
$\mu = NV_{\slm C}$ and $\mu^* = NV_{\slm C} (E_1)$, the coupling is
renormalized as in eq.~\ee{running},
\eq
\mu^* = \frac{\mu}{1 + \frac{\mu}{2} \ln(M / m)}. \e{mustar}
\eqe
The coupling $g$ is not renormalized, by Migdal's theorem.
At $E_1$, scaling has brought the dimensionless magnitude of $g$ to order~1.
Integrating out the phonons produces a new $O(1)$ contribution $V_{\slm p}$
to the four-Fermi interaction.  It is conventional to define $N V_{\slm p}
= - \lambda$, so the total four-Fermi interaction just below $E_1$
is
\eq
NV(E_1^-) = \mu^* - \lambda.
\eqe
What happens next depends on the sign of $\mu^* - \lambda$.  If it is positive,
then $V(E)$ below $E_1$ just grows weaker and weaker---not very exciting.
If, however, it is negative, then the coupling grows and becomes strong at a
scale
\eq
E_c = E_1 e^{ - 1/( \lambda - \mu^*) } = E_0 \left( \frac{m}{M}
\right)^{1/2} e^{ - 1/( \lambda - \mu^*) }.  \e{crit}
\eqe
What happens at strong coupling?  It seems to be a fairly general rule of
nature that gapless fermions with a strongly attractive interaction are
unstable, so that a fermion bilinear condenses, breaks symmetries, and produces
a gap.  In QCD this breaks the chiral symmetry.  Here, the attractive channel
involves two electrons, so the condensate breaks the electromagnetic $U(1)$
and produces superconductivity: this is the BCS theory.
Because of the simplicity of Fermi liquid theory, it is not
necessary to guess about the condensation.  Calculating the quantum effective 
potential
for the electron-electron condensate, interactions where a pair of electrons
vanish into the vacuum are marginal because the pair has zero momentum.  The
one-loop graph is thus marginal, but all higher graphs are irrelevant.
In other words, the effective potential sums up the
`cactus' graphs, the same as in large-$N$ $O(N)$
models and mean field theory.  The gap and the critical temperature are
indeed of the form~\ee{crit}, with calculable numerical coefficients.

So BCS superconductivity is another example of `a marginal coupling grows
strong and something interesting happens.'  The simple renormalization group
analysis gives a great deal of information.  It does not get the $O(1)$
coefficient in the critical temperature, but it gets something else which is
often omitted in simple treatments of BCS: the renormalized Coulomb repulsion.
This correction is significant for at least two reasons.  The first is the 
likelihood of superconductivity, which depends on $\mu^* - \lambda$ being
negative.  The initial four-Fermi interaction, being a screened Coulomb
interaction, is most likely to be repulsive, positive.  The phonon contribution 
$V_{\slm p}$ is attractive, negative, because it arises from second order
perturbation theory (hence the sign in the definition of $\lambda$).  Now,
$\mu^*$ and $\lambda$ are both of order~1; since the only small parameter is
$m/M$, this just means that they do not go as a power of $M$ in the $M \to
\infty$ limit.  In fact, they are both generally within an order of magnitude
of unity, and there is a simple model of solids
in which they are equal.\footnote{
See chapter~26 of Ashcroft and Mermin.  It might seem surprising that
the phonon interaction, which vanishes when $M \to \infty$, can compete with
the Coulomb interaction, which does not.  The point is that this is only true
at energies below the Debye scale, which also vanishes as $M \to \infty$.
At these low energies, the $M^{-1}$ from the interaction cancels against an
$M^{-1}$ in the denominator of the phonon propagator. }  This model, however,
does not take into account the renormalization~\ee{mustar}.  The
renormalization is substantial because the logarithm is approximately 10,
and one sees that $\mu^*$ cannot exceed $0.2$ no matter how large $\mu$ is.
Thus, superconductivity is more common than it would otherwise be.

The renormalization of the Coulomb correction is also important to the
isotope effect, the variation of the critical temperature with ion mass $M$.
As $M$ is varied, there is an overall $M^{-1/2}$ in the critical
scale~\ee{crit}, coming from the change in the Debye scale.  There is also
an implicit dependence on $M$ in $\mu^*$.  When the Debye scale is
lowered, the Coulomb interaction suffers 
more renormalization and so is reduced;
this goes in the opposite direction, favoring superconductivity.  The
exponent
\eq
\alpha = - M \del_M E_c = \frac{1}{2} \Biggl\{ 1 - \frac{\mu^{*2}}
{(\lambda - \mu^*)^2} \Biggr\}
\eqe
is the naive $\frac{1}{2}$ when $\mu^*$ is much less than $\lambda$,
but as $\mu^* / \lambda$ increases $\alpha$ can be substantially smaller, as
is found in some materials.

When the coupling $\lambda - \mu^*$ is large, 
the renormalization group analysis
does not give a large ratio between $E_1$ and $E_c$.  It is then not possible
simply to integrate the phonons out; the full phonon 
propagator must be retained.
This leads to the Eliashberg theory, which is still solvable in the sense that
it can be reduced to an integral equation.  Because of Migdal's theorem, the
Schwinger-Dyson equation for the two-point function closes.  This resolves the
last of the puzzles with which this lecture began.  The Eliashberg theory
involves several phenomenological functions, which are precisely those
appearing in the effective action.

Now, what about high $T_c$?  Figure~6 shows a graph of resistivity versus
temperature for a typical high-$T_c$ material.  
\begin{figure}
\begin{center}
\leavevmode
\epsfbox{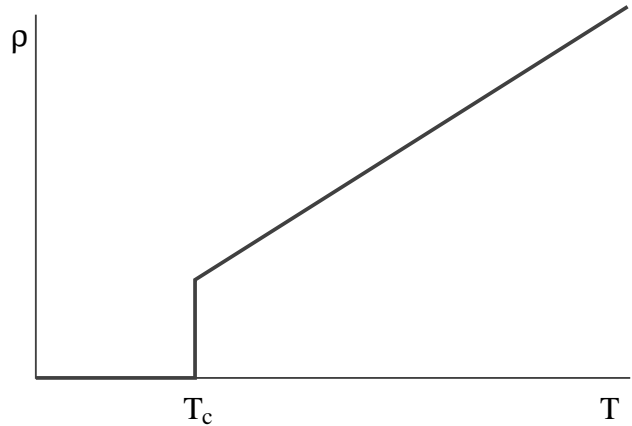}
\end{center}
\caption[]{Resistivity versus temperature in a typical high-$T_c$ material:
zero below $T_c$, and linear above.}
\end{figure}
One sees the expected drop to
zero at low temperature, but there is also something very puzzling:
the resistivity is linear to good accuracy above $T_c$,
\eq
\rho(T) \sim A + B T.  \e{linres}
\eqe
By comparison, the resistivity above $T_c$ in an ordinary metallic
superconductor goes as
\eq
\rho(T) \sim A + C T^5. \e{ordres}
\eqe
How does this relate to what we have learned?  We know that conductors
are very simple, nearly free, and that any physical effect will have some
definite energy dependence governed by the lowest dimension operator that could
be responsible.  For example the $T^0$ resistivity is from impurity
scattering.\footnote
{Incidentally, there is perhaps some indication that $A$ is anomalously small,
even zero, in the best-prepared high-$T_c$ materials.}
The $T^5$ resistivity is from phonon scattering; the high power of temperature
is because we are below the Debye temperature, so only the long-wavelength
phonons remain, their contribution suppressed by phase space and the $\bq$ in
the vertex.  What can give $T^1$?  {\it Nothing.} 
Write down the most general possible effective Lagrangian and 
there is no operator or process that would this power of the temperature.  This
is one of several related anomalies in these materials.  To
steal a phrase from Mike Turner, figure~6 shows the conductor from Hell.

To be precise, there is nothing of this magnitude in the generic Fermi
liquid theory, but in special cases the infrared divergences are enhanced and
new effects are possible.  For example, consider free electrons on a square
lattice of side $a$, with amplitude $t$ per unit time to hop to one of the
nearest neighbor sites.  This models the CuO planes of the high-$T_c$
materials.  Going to momentum space, the one-electron energy works out to
\eq
\varepsilon (\bp) = - t (\cos p_x a + \cos p_y a).
\eqe
For half-filling, $\varepsilon_{\slm F} = 0$, the Fermi sea is as shown in
figure~7.  
\begin{figure}
\begin{center}
\leavevmode
\epsfbox{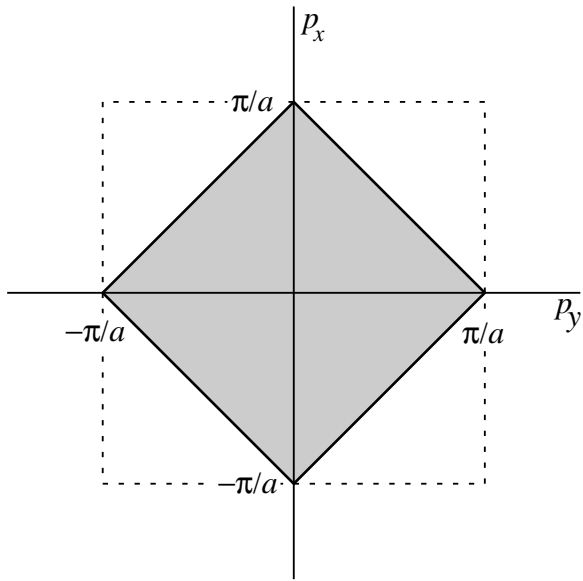}
\end{center}
\caption[]{Diamond-shaped Fermi surface for the half-filled squared lattice.
The dashed lines at $|P_{x,y}| = \pi/a$ bound the Brillouin zone and are
periodically identified.}
\end{figure}
There are two special features.  The first is the presence of 
{\it van Hove} singularities, the corners of the diamond where $\del_\bp
\varepsilon = \bv_{\slm F}$ vanishes.  At a van Hove singularity the
density of states $N$ diverges logarithmically, enhancing the interactions.
The second special feature is {\it nesting}, which means that the opposite
edges of the Fermi surface differ by a fixed translation $(\pi/a, \pi/a)$.
Because of nesting, the interaction
between an electron-hole pair with total momentum 
$(p_x,p_y) = (\pi/a, \pi/a)$ becomes
marginal just as for an electron pair of zero momentum.  For positive
$V$ this is attractive and favors condensation, producing either a
position-dependent charge density $\psi_\sigma^\dagger \psi_\sigma$ (a charge
density wave), or a position-dependent spin density $\psi_\sigma^\dagger
\mbox{ \boldmath $\sigma$ }\!\!_{\sigma \sigma'} \psi_{\sigma'}$ 
(antiferromagnetism).  So here are
two more phenomena that can arise from the growth of a marginal coupling.
It has been proposed that a Fermi surface which has van Hove singularities,
or which is nested, or which sits near the antiferromagnetic transition,
would have sufficiently enhanced infrared fluctations to account for the
anomalous behavior of the high-$T_c$ materials.

Is this plausible?  Recall our earlier observation that the shape of the Fermi
surface is a relevant parameter---a shift in the Fermi surface acts like a mass,
cutting off the enhanced infrared fluctuations.
For these to persist down to low energy the Fermi surface must 
be highly fine-tuned.\footnote
{Notice that the van Hove singularity is less unnatural than nesting, since
the former requires only a single parameter to be tuned (the level, which must
pass through the point where $\bv_{\slm F} = 0$), while a very large (in 
principle infinite) number must be tuned for nesting.}
This might occur in a few very particular
substances, but can it be happening here? The linear resistivity is present
in many different high-$T_c$ materials 
(though not all), and it is stable against
changes in the doping (filling fraction) of order 
5 to 10 percent.  The width of
the electron band ($E_0 \sim 4t$) is roughly 2 eV, 
so this represents a shift in
the Fermi surface of order $0.1$ eV.  On the other hand, the anomalous behavior
persists below 100 K (0.01 eV).  In one substance,
Bi2201, which is in the high-$T_c$ class although its transition
temperature is rather low, the resistivity 
is beautifully linear from 700 K down
to 7 K ($<0.001$ eV), and is stable against changes in doping.  
So most of the high-$T_c$ materials must be fine-tuned to accuracy
$10^{-2}$ and Bi2201 must be fine-tuned to accuracy $10^{-3}$.  This is not
obviously bad, since there are thousands of substances from which to choose.
But if fine tuning is the answer, one would expect it to be spoiled by
a small change in the doping, which will shift all the relevant parameters,
and this is not the case.
In Bi2201, in particular, a fine tuning to accuracy better than $10^{-3}$
would have to survive shifts of order $10^{-1} E_0$ in the
relevant parameters.
This seems inconceivable
unless there are no relevant parameters at all.  The special
Fermi surface cannot account for the anomalous behavior.  We must find a
low-energy effective theory which is {\it natural}, in the same sense as used
in particle physics: there are no relevant terms allowed by the symmetries.

We should mention one other possibility.  At temperatures above their
frequency, phonons do give a linear resistivity.  In the high-$T_c$ materials,
the phonon spectrum runs up to $0.05$ eV, with only 5 to 10 per cent of the
density of states below 0.01 eV.  
To account for the resistivity, even excluding
2201, one would have to suppose that this small fraction accounts for almost
all the resistivity, which is implausible.  And only the very
longest-wavelength phonons, which give the $T^5$ behavior, survive down to
the 0.001 eV of 2201.

It appears that the low energy excitations are not described by the effective
field theory that we have described, but by something different.  Perhaps
we should not be surprised by this, since we began with a guess about the
spectrum.  Rather, it is surprising that the guess is correct in so many
cases.\footnote{ Though there are some other examples of apparent non-Fermi
liquid behavior. }  From studies of strongly interacting electron systems one
can motivate several other guesses.\footnote{ The book by Fradkin gives a
review of recent ideas. }  Typically the low energy theory includes fermions
and also gauge fields (which seem like a good thing from the point of view of
naturalness) and/or scalars (which do not).   For example, anyon theories can
be regarded as 
fermions interacting with a gauge field which has a $T$-violating
Chern-Simons action.  Another possibility is fermions with a
$T$-preserving Maxwell action.  The normal-state properties of the anyon theory
do not seem to have been studied extensively.  The $T$-preserving theory has
been argued to give the right behavior, but it is strongly coupled and not
well understood.  Anderson has proposed what is apparently the Fermi-liquid
theory but with the four-Fermi interaction singular in momentum space.  The
effect of the singularity is to enhance the infrared behavior so that the
system
behaves as though it were one-dimensional (which, as we have noted, is always
marginal).  It is difficult to understand the origin of this interaction; in
particular, it is long-ranged in position space but is not mediated by any
field in the low energy theory, which seems to violate locality.  Finally,
there is a semi-phenomenological idea known as the `marginal Fermi liquid,'
which I have not been able to translate into effective field theory language.

Notice that we have not discussed the mechanism for superconductivity itself;
the normal state is puzzling enough.
If one can figure out
what the low energy theory is, the mechanism of condensation will presumably be
evident.

In terms of sheer numbers, there seems to be a move away from
exotic field ideas and back to more conventional ones in this subject.
This is largely because none of the new theories has made the sort of clear-cut
and testable predictions that the BCS theory does.  From my discussions with
various people and reading of the literature, however, it seems 
that attempts to explain the normal state properties in a conventional way
always require the extreme fine-tunings described above.
This seems to be a subject where particle theorists can contribute: the basic
issue is one of field theory, where many of the unfamiliar details of
solid-state physics are irrelevant (in the technical and colloquial senses).

{\bf Q:}  So `quasiparticle' means the quantum of an effective field?

{\bf A:}  More-or-less.  As used in condensed matter physics, the term has
one additional implication that will not always hold in effective field theory:
that the decay rate of the quasiparticle vanishes faster 
than the energy $E$ as $E$ goes to zero.  There may be systems for which
this is not true (a nonrelativistic system at a nontrivial fixed point being the
obvious case), but where one still expects the low energy fluctuations to
be represented by some field theory.

{\bf Exercise:} Consider the term 
\eq
\int dt\, d^3 \bp_1 d^3 \bp_2 \, U(\bp_1, \bp_2) \psi_\sigma^\dagger(\bp_1)
\psi_\sigma(\bp_2).
\eqe
This is impurity scattering: notice the lack of momentum conservation.
Show that this is marginal, and that its beta-function vanishes.

{\bf Exercise:} Now consider an impurity of spin $s$,
which can exchange spin with the electron:
\eq
\int dt\, d^3 \bp_1 d^3 \bp_2 \, J(\bp_1, \bp_2) \psi_\sigma^\dagger(\bp_1)
\mbox{ \boldmath $\sigma$ }\!\!_{\sigma\sigma'} 
\psi_{\sigma'}(\bp_2) \cdot {\bf S},
\eqe
where $\bf S$ are the spin-$s$ matrices for the impurity.  Show that
this is marginal and that the beta-function is negative, taking $J$ to be
a constant for simplicity.  This
is the Kondo problem. The nonvanishing beta-function means that the coupling
grows with decreasing energy (for $J$ positive). This is vividly seen in
measurements of resistivity as a function 
of temperature, which increases as $T$
decreases rather than showing the simple constant 
behavior of potential scattering. 
When the coupling gets strong, a number of behaviors are possible,
depending on the value of $s$, sign of $J$, and various generalizations.
In particular, in some cases one finds fixed points with critical behavior
given by rather nontrivial conformal field theories: more examples of the
interesting things that can happen when a marginal coupling gets strong!

{\bf Exercise:} Show that if the Fermi surface is right at a van Hove
singularity, then under scaling of the energy to zero and of
the momenta {\it toward the singular point,} the four-Fermi interaction is
marginal in {\it two} space dimensions.
In other words, if all electron momenta in a graph lie near the
singularity, the graph is marginal: one does not have the usual simplifications
of Landau theory.

\subsection*{Acknowledgements}

I would like to thank I. Affleck, B. Blok, 
N. Bulut, F. de Wette, V. Kaplunov\-sky,
A. Ludwig, M. Marder, 
J. Markert, D. Minic, M. Natsuume, D. Scalapino, E. Smith,
and L. Susskind for helpful remarks and
conversations.  I would also like to thank R. Shankar for discussions of his
work and S. Weinberg for comments on the manuscript.
This work was supported in part by National Science Foundation
grants PHY89-04035 and PHY90-09850, by the Robert A. Welch Foundation,
and by the Texas Advanced Research Foundation.

\section*{Bibliography}
\parskip=6pt
\baselineskip=18pt

\subsection*{Lecture 1}

Wilson's approach to the effective action is developed in\\[4pt]
\hspace*{.5in} K. G. Wilson, Phys. Rev. {\bf B4} (1971) 3174, 3184.\\[4pt]
For further developments see\\[4pt]
\hspace*{.5in} K. G. Wilson and J. G. Kogut, Phys. Rep. {\bf 12} (1974) 75;\\
\hspace*{.5in} F. J. Wegner, in {\it Phase Transitions and Critical Phenomena,
Vol. 6,} ed.~C. Domb and M. S. Green, Academic Press, London, 1976;\\
\hspace*{.5in} L. P. Kadanoff, Rev. Mod. Phys. {\bf 49} (1977) 267;\\
\hspace*{.5in} K. G. Wilson, Rev. Mod. Phys. {\bf 55} (1983) 583.\\[4pt]
For the treatment of perturbative renormalization from Wilson's point
of view see\\[4pt]
\hspace*{.5in} J. Polchinski, Nucl. Phys. {\bf B231} (1984) 269;\\
\hspace*{.5in} G. Gallavotti, Rev. Mod. Phys. {\bf 57} (1985) 471.\\[4pt]
As I hope is clear from the discussion, these ideas do not depend on
perturbation theory, and have been used to prove the existence of the 
continuum limit nonpertubatively in asymptotically free theories.  This
is done for the Gross-Neveu model in\\[4pt]
\hspace*{.5in} K. Gawedzki and A. Kupiainen, Comm. Math. Phys. {\bf 102} (1985)
1,\\[4pt]
and for $D=4$ 
non-Abelian gauge theories in a series of papers culminating in\\[4pt]
\hspace*{.5in} T. Balaban, Comm. Math. Phys. {\bf 122} (1989) 355.

The idea that pion physics can be encoded in a Lagrangian appeared in\\[4pt]
\hspace*{.5in} S. Weinberg, Phys. Rev. Lett. {\bf 18} (1967) 188;
Phys. Rev. {\bf 166} (1968) 1568,\\[4pt]
and was developed further in\\[4pt]
\hspace*{.5in} S. Coleman, J. Wess, and B. Zumino, 
Phys. Rev. {\bf 177} (1969) 2239;\\
\hspace*{.5in} C. G. Callan, S. Coleman, J. Wess, and B. Zumino, Phys. Rev. 
{\bf 177} (1969) 2247;\\
\hspace*{.5in} S. Weinberg, Physica {\bf 96A} (1979) 327;\\
\hspace*{.5in} J. Gasser and H. Leutwyler, Ann. 
Phys. (N.Y.) {\bf 158} (1984) 142;\\
\hspace*{.5in} H. Georgi, {\it Weak Interactions and Modern Particle Theory,}
Benjamin/Cummings, Menlo Park, 1984.

Two other classic papers are\\[4pt]
\hspace*{.5in} T. Appelquist and J. Carazzone, Phys. Rev. 
{\bf D11} (1975) 2856;\\
\hspace*{.5in} S. Weinberg, Phys. Lett. {\bf B91} (1980) 51.\\[4pt]
A recent application, to the `oblique' corrections in
weak interaction theory, is\\[4pt]
\hspace*{.5in} H. Georgi, Nucl. Phys. {\bf B361} (1991) 339.

An interesting situation arises when heavy particles are in interaction with
light, with all kinetic energies small compared to the heavy particle
rest masses.  By assumption, the number of heavy particles is then fixed,
and the low energy degrees of freedom consist of the light {fields} plus
the positions and the spin and internal quantum numbers of the heavy {
particles}.  The case of heavy and light quarks has recently been studied
extensively; for some discussions see\\[4pt]
\hspace*{.5in} H. D. Politzer and M. B. Wise, 
Phys. Lett. {\bf B208} (1988) 504;\\
\hspace*{.5in} H. Georgi, Phys. Lett. {\bf B240} (1990) 447.\\[4pt]
The case of pions in interaction with slow nuclei has been studied in\\[4pt]
\hspace*{.5in} S. Weinberg, Nucl. Phys. {\bf B363} (1991) 3.

For another introduction to effective field theory, see\\[4pt]
\hspace*{.5in} G. P. Lepage, in Proceedings of TASI 1989.\\[4pt]
An interesting history of the changing philosophy of renormalization is\\[4pt]
\hspace*{.5in} T. Y. Cao and S. S. Schweber, {\it The Conceptual Foundations
and Philosophical Aspects of Renormalization Theory,} 1991 (unpublished).\\[4pt]
This is a rare instance where the historians are ahead of most textbooks!

Obviously this is a very selective list.  In particular, I have chosen papers
based on pedagogic value rather than priority,
taking those that apply the effective Lagrangian language in a general way.

\subsection*{Lecture 2}

The Wilsonian approach to Fermi liquid theory is developed in\\[4pt]
\hspace*{.5in} G. Benfatto and G. Gallavotti, 
J. Stat. Phys. {\bf 59} (1990) 541;
Phys. Rev. {\bf B42} (1990) 9967;\\
\hspace*{.5in} R. Shankar, Physica {\bf A177} (1991) 530.\\[4pt]
It is developed further in unpublished work of Shankar, who likens the Landau
theory to a large-$N$ expansion, with $N$ corresponding to the area of the
Fermi surface (in units of $E$).  See also\\[4pt]
\hspace*{.5in} P. W. Anderson, {\it Basic Notions of Condensed Matter Physics,}
Benjamin/Cummings, Menlo Park, 1984.

Standard treatments 
of Fermi-liquid theory and superconductivity can be found in\\[4pt]
\hspace*{.5in} A. A. Abrikosov, L. P. Gorkov, and  
I. E. Dzyaloshinski, {\it Methods of Quantum Field Theory in Statistical
Mechanics,} Dover, New York, 1963;\\
\hspace*{.5in} P. Nozi\'eres, 
{\it Theory of Interacting Fermi Systems,} Benjamin, New York,
1964;\\
\hspace*{.5in} J. R. Schrieffer, {\it Theory of Superconductivity,} 
Benjamin/Cummings, Menlo Park, 1964;\\
\hspace*{.5in} G. Baym and C. Pethick, {\it Landau Fermi-Liquid 
Theory,} Wiley, New York, 1991.\\[4pt]
Baym and Pethick is primarily concerned with
liquid $^3$He, another example of a Fermi liquid.
Superconductivity at strong coupling 
is discussed in Abrikosov, et.~al., section~35,
and Schrieffer, section~7-3.  The renormalization of the Coulomb interaction
is discussed in\\[4pt]
\hspace*{.5in} P. Morel and P. W. Anderson, 
Phys. Rev. {\bf 125} (1962) 1263.\\[4pt]
A thorough introduction to solid state physics can be found in\\[4pt]
\hspace*{.5in} N. W. Ashcroft and N. D. Mermin, {\it Solid State Physics,}
Holt, Rinehart and Winston, New York, 1976.

The anomalous properties of the normal state in the high $T_c$ materials
are discussed by Philip Anderson and Patrick Lee in\\[4pt]
\hspace*{.5in} {\it High Temperature Superconductivity, Proceedings of the 
1989 Los Alamos
Symposium,} ed.~K. S. Bedell, et.~al., Addison-Wesley, Redwood City CA,
1990.\\[4pt]
See also the interesting discussion between Anderson and Schrieffer in\\[4pt]
\hspace*{.5in} Physics Today (June 1991) 55.\\[4pt]
The proceedings of the meeting High Temperature Superconductors III, 
Kanazawa, Japan, 1991, give
a comprehensive overview of recent theoretical ideas and experimental results.
They are published as\\[4pt]
\hspace*{.5in} Physica, {\bf C185-C189}.\\[4pt]
For see a nice plot of the linear resistivity in Ba2201, see\\[4pt]
\hspace*{.5in} S. Marten, A. T. Fiory, 
R. M. Fleming, L. F. Schneemeyer, and J. V.
Waszczak, Phys. Rev. {\bf B41}, (1990) 846.\\[4pt]
For some Ba2201 data which shows the stability of the linear resistivity
against changes in doping, see\\[4pt]
\hspace*{.5in} P. V. Sastry, J. V. Yakhmi, R. M. Iyer, C. K. Subramanian,
and R. Srinivasan, Physica {\bf C178} (1991) 110.

A review of recent work on strongly coupled electron systems is\\[4pt]
\hspace*{.5in} E. Fradkin, {\it Field Theories of Condensed Matter Systems,}
Addison-Wesley, Redwood City, 1991.

\end{document}